\documentclass[reprint,superscriptaddress,amsmath,amssymb]{revtex4-1}
\usepackage{xcolor}
\usepackage{graphicx} 

\newcommand{\nupos}{\nu_{\text{{p}}}}
\newcommand{\nuneg}{\nu_{\text{{e}}}}

\newcommand{\rpos}{r_{\text{{p}}}}
\newcommand{\rneg}{r_{\text{{e}}}}

\newcommand{\Npos}{N_{\text{{p}}}}
\newcommand{\Nneg}{N_{\text{{e}}}}
\newcommand{\Nnuc}{N_{\text{{n}}}}

\newcommand{\ppos}{p_{\text{{p}}}}
\newcommand{\pneg}{p_{\text{{e}}}}

\newcommand{\cQ}{\mathcal{Q}}

\newcommand{\alphaS}{\alpha_{\text{\tiny{S}}}}

\newcommand{\veps}{\varepsilon}

\newcommand{\Ddr}{\frac{d\phantom{s}}{d{r}}}

\newcommand{\mEL}{m_{\text{e}}}
\newcommand{\mPR}{m_{\text{p}}}

\newcommand{\vect}[1] {\boldsymbol{{ #1}} }

\newcommand{\Rset}{\mathbb{R}}

\newcommand{\pV}{{\vect{p}}}            
\newcommand{\qV}{{\vect{q}}}            
\newcommand{\sV}{{\vect{s}}}            
\renewcommand{\leq}{\leqslant}
\renewcommand{\geq}{\geqslant}

\begin{document}
	
\title{How much electric surcharge fits on ... a ``white dwarf'' star?\vspace{-10pt}}
	
\author{Parker Hund$^1$ and Michael K.-H. Kiessling}

\affiliation{Department of Mathematics, Rutgers University,
                110 Frelinghuysen Rd., Piscataway, NJ 08854, USA}\vspace{-10pt}
\email{\vspace{-25pt}ph325@math.rutgers.edu}\email{\vspace{-25pt}miki@math.rutgers.edu}

\begin{abstract}
\noindent 
 The question how much surplus of electric charge (``surcharge'')
fits on an object is generally very difficult to answer. 
 Here it is shown that it is easy to answer when the object is a \emph{failed white dwarf} star 
(a brown dwarf in its ground state) made of protons and electrons:
Given the number of protons, how many electrons can there be?
 Surprisingly, the answer (in the form: as few as $\mathfrak{A}$ and as many as $\mathfrak{B}$) 
is independent of the speed of light $c$ and the Planck quantum $h$, even when the star is stabilized
against collapse by relativistic quantum mechanics.
\end{abstract}
$\phantom{xi}$\hfill 1 \vspace{-9pt}
\maketitle

\section{Introduction}\vspace{-10pt}

 Everyday matter typically appears to be electrically neutral, yet most people
are familiar with the fact that objects of all kinds can sometimes be charged with 
a surplus of electricity, our bodies included. 
 Anyone who has ever walked on synthetic carpet with the wrong kind of shoes 
and then touched a door handle knows how unpleasant the ensuing spark can be
as the accumulated surplus of charge, or surcharge for short, is neutralized.

 One could ask, how much electrical surcharge can a general object hold? 
 Empirically the answer is: never much.
 But how much, exactly, depends on many things, such as on the material which is
being charged, on its shape, its temperature, the environment, and such. 

 Take the familiar example of the high school physics experiment where in a dry-air class room
a glass rod is rubbed with a silk cloth and then held near to the head of a volunteer student.
 The student's hair promptly will be attracted by the glass rod and easily made to stand up. 
 This experiment is often used to introduce students to the notion of electric charge and 
electric force, for its entertainment effect. 
 As for the dependence of the outcome on the environment:
Do the experiment when the weather (and the classroom) is extremely humid and no matter how 
hard you rub the rod, the effect will not be very pronounced. 

 Back to our question (in this example): How much {surcharge} can the rod hold?
 In some more advanced classes the teacher will take the rod to a Coulomb meter after rubbing it, which
reveals that the rod carried a {surplus of positive charge}. 
 Since positive charge is localized in the barely movable nuclei of the atoms, it means that some of the 
negatively charged electrons of the atoms'  electronic hulls have been stripped off of the glass 
rod by the silk cloth. 
 One can repeat this many times and will note that the amount of charge imbalance on the glass rod will stay 
below some upper limit, no matter how hard and long you rub it with the cloth. 
  Unfortunately, while easy to demonstrate, the glass-rod-rubbed-by-a-silk-cloth effect is not easy to calculate.

 Incidentally, the answer to our question should not be confused with (self-)capacitance, a 
superficially related concept treated in introductory E\&M courses, e.g. \cite{HWR}, \cite{Purcell}; see
in particular \cite{JacksonBOOKb}, chpt.1, sect.11.
 In this case one considers a perfect conductor held at a constant electric potential against ground, and defines its
capacitance as the ratio of the charge $Q=\cQ(V)$ it holds to the voltage $V$ which is applied.
 However, the ratio $Q/V$ yields only the slope of the response function $\cQ(V)$ in its linear regime. 
 Eventually $\cQ(V)$ becomes nonlinear and reaches either a positive maximum or negative minimum, depending on
whether electrons are being stripped off of the conductor or transferred to it.
 Our question of how much electric surcharge fits onto an object refers to these two extremal 
amounts of electrons that can be stripped off, respectively deposited on an object.
 
 Moreover, ``object'' may or may not mean a conductor (think of the glass rod).
 It simply is a physical system consisting of a fixed number $N_+^{}$ of nuclei 
arranged in an (essentially) fixed shape, {comprising a total of $\Npos^{\mbox{\tiny{tot}}}\geq N_+$ protons, 
$\Npos$ of which are free (hydrogen nuclei) and the rest is bound in more massive nuclei.}
 For a neutral object, the number of electrons {$\Nneg=\Npos^{\mbox{\tiny{tot}}}$}.
 Since the electron mass $\mEL$ is so much smaller than the proton mass $\mPR$,
and since empirically any possible surcharge is small, the mass of the object is 
essentially determined by the $N_+$ nuclei.

 Furthermore, an object does not need to be macroscopic. 
 An important example is an atomic ion, having a single nucleus ($N_+=1$) with 
$\Npos^{\mbox{\tiny{tot}}}=Z\in\{1,...,118\}$ elementary charges $e$.
 There are positive and negative ions.
 Empirically it becomes exceedingly difficult with growing $Z$ to strip off all $Z$ electrons from a neutral atom, 
though in principle it can be done. 
 Yet, and still empirically, not more than two or three excess electrons can be
placed on a neutral atom to create a negative ion, and an $\alpha$ particle (a ${}^4_2$He nucleus with two elementary
charges) does not seem to bind more than two electrons.
 Theoretically, a point nucleus and no electrons is a quantum-mechanically trivial problem. 
 In the exactly solvable non-relativistic hydrogenic problem a nucleus with $Z\geq 1$ elementary charges
always binds one electron.
 Abstractly it has been proved that the $\Nneg$-electron Schr\"odinger--Pauli Hamiltonian of an atomic ion 
with a nucleus of $Z$ elementary charges always has a bound state if $1\leq \Nneg\leq Z$, see \cite{LiebSeiringer}.
 However, it still is an open problem to theoretically determine the number 
of excess electrons that such a nucleus can bind, as per the many-body Schr\"odinger--Pauli equation;
see \cite{LiebSeiringer}, {pp.164--178.}

 In this paper we show that the answer to our question can be found explicitly for what we call
a failed white dwarf star; see below for a definition.
 The study of theoretical models of white dwarfs has already been emphasized as 
being an excellent way for an undergraduate to connect various topics in physics; cf. \cite{SilbarReddy} and \cite{Garfinkle}. 
 We re-emphasize this by drawing attention to yet another topic, that of the maximal excess charge, a difficult 
open research question for atomic ions \cite{LiebSeiringer}, yet 
 easy to answer in stellar models of the type discussed in \cite{KNY} and \cite{RRb}.

 While the theory of white dwarfs goes back to Chandrasekhar's pioneering work \cite{Chandra},
far from being sorted out long ago in all generality, this is still an active area of research, 
including the question of its overall charge density \cite{KNY}, \cite{RRb}.
 Moreover, white dwarf stars can feature strong magnetic fields; \cite{DM1}, \cite{DM2}, \cite{BB}, \cite{CFC}, \cite{CFD}. 
 Magnetic fields, however, complicate the equations considerably, {as does rotation}, 
Wigner crystalization, finite temperature corrections, and so on. 
 Since our question about the electric excess charge seems not to have been answered 
even for the most basic non-magnetized, {non-rotating}, zero-temperature white dwarf models as used in
\cite{KNY} and \cite{RRb}, we will only consider similarly idealized stellar models in our paper.
 Our study should not be confused with an attempt to explain observed astrophysical effects.

 Indeed, we reduce the model of \cite{KNY} even further.
 In \cite{KNY}, the complicated mixture of nuclei species in a real white dwarf, which
in addition to protons ${}^1_1$H includes $\alpha$ particles ${}^4_2$He, carbon nuclei 
${}^{12}_{\;6}$C, oxygen nuclei ${}^{16}_{\;8}$O, etc., is jointly represented by a single
positive effective density function, somewhat in the spirit of Chandrasekhar's original work \cite{Chandra}. 
 The protons and electrons are fermions; the other nuclei mentioned above are bosons; yet in \cite{KNY} 
both the electrons and the effective mix of nuclei are treated as completely degenerate ideal Fermi gases.
 To avoid a discussion whether or not this is justified, we consider the special case of the model of \cite{KNY} 
with only two species of fermions: individual protons and electrons.

 This means such a star never ignited (otherwise a certain percentage of the protons would have undergone
nuclear fusion into $\alpha$ particles, to say the least).
 Put differently, the star failed to ignite, and so we refer to this as a failed star.
 This yields an estimated upper bound of roughly $9\cdot 10^{55}$ protons for such an object (more 
and the central particle density would cross the threshold for sustained nuclear fusion of protons into $\alpha$ particles).
 On the other hand, one also needs more than roughly $1.5\cdot 10^{55}$ protons (amounting to about 13 Jupiter masses) to
speak of a star, not a planet --- clearly, this borderline is not a law of nature but depends on
the consensus of the astrophysical community.
 The closest one gets in terms of real objects in space are stars more commonly known as brown dwarfs \cite{Jill}, 
although a first generation brown dwarf would have about $92\%$ protons and $8\%$ $\alpha$ particles,
and observed brown dwarfs have not yet cooled down enough to near to their lowest energy state.
 The zero-temperature approximation on the other hand has been estimated to be quite accurate for the purpose of computing 
the overall density structure of a white dwarf, so we prefer the slightly more informative ``failed white dwarf'' 
for the idealized kind of objects we discuss.

 A non-relativistic theory suffices to describe failed white dwarfs.
 The only forces at work are Newton's gravity, Coulomb's electricity, and the quantum-mechani\-cal stabilization
against collapse, in particular the Pauli principle, modelled by gradients of the non-relativistic degeneracy 
pressures of two ideal Fermi gases.

 Although relativistic effects are quantitatively unimportant for low-mass stars,
a Chandrasekhar-type special-relativistic treatment of the quantum-mechanical pressure-density relation is 
readily accomodated, as we shall explain.
 We will also briefly comment on the more complicated general relativistic model of the kind presented in \cite{RRb} 
(without going into any technical details).

 Once the model equations are formulated, it is remarkably easy to give a compelling 
derivation of the allowed excess charge.
 Aside from the simplicity with which the result is determined,
the perhaps most surprising outcome of our study is the fact that the permitted number of electrons per proton, $\Nneg/\Npos$, 
in a failed white dwarf star made of electrons and protons
neither depends on the Planck quantum $h$ nor on the speed of light $c$, even though
it is quantum mechanics which prevents these objects from collapsing so long as their mass stays sufficiently below the 
Chandrasekar critical mass $\propto(\hbar c/G)^\frac32/\mPR^2$ beyond which special-relativistic effects destabilize
the quantum-mechanical stabilization (general relativity destabilizes even more).
 This suggests that the same bounds on $\Nneg/\Npos$ could be obtained with purely Newtonian 
computations, involving only the electrical Coulomb and gravitational Newton forces. 
 Indeed, this is the case, yet this Newtonian argument gets the structure of the ground state entirely wrong!
 How such a seemingly paradoxical situation can be understood is explained in our paper. 
 We also explain why general relativity will not likely change the allowed $\Nneg/\Npos$ interval.

 The rest of the paper is structured as follows:

  First, in section II, we show with very elementary arguments
 that purely Newtonian physics yields a completely collapsed ground state of many protons and electrons,
as long as $\Nneg/\Npos$ lies in a certain interval.
 Obviously this interval has to be independent of $\hbar$ and $c$.
 This already is the correct answer to our surcharge question, but the model cannot possibly
explain why this answer remains correct also in quantum mechanics and relativity theory.

 In section III we recall that quantum mechanics 
prevents such a non-relativistic system from collapsing to a point,
 cf. \cite{LiebSeiringer}, and we give an elementary explanation how the Pauli principle accounts for a spread out 
stellar ground state.
 Our surcharge question now takes a sharp mathematical form in terms of the existence
of a lowest eigenvalue of the Hamiltonian $H$, in analogy to the question about ions mentioned above.
 We then explain that the large number of particles in our problem allows one to answer the question in the 
context of the continuum approximation, known as  Thomas--Fermi theory.

 In section IV we state the Thomas--Fermi type equations of a failed white dwarf star, 
which are simply the \emph{classical} Euler equations of mechanical equilibrium of two 
charged fluids, yet with \emph{quantum-mechanical} pressure functions obtained for two ideal Fermi gases 
of spin-$\frac12$ particles; cf. \cite{KNY}.
 Compared to the usual local neutrality approximation, which yields a second-order ordinary differential
equation for a single-density model polytrope of index $n=3/2$, \cite{Emden}, \cite{Chandra}, \cite{KippenhahnWeigert}, 
the two-species model furnishes a coupled system of two  equations with polytropic index $n=3/2$. 
 Interestingly, as we will show, the scaled $n=3/2$ standard polytrope also furnishes one distinguished 
exact solution pair of the two-species model, which 
facilitates the comparison with the models discussed in  \cite{Emden}, \cite{Chandra}, \cite{KippenhahnWeigert}, 
and \cite{SilbarReddy}, \cite{Simon}, \cite{Pesnell}, \cite{GjerlovPesnell}. 
 As is clear from some of these cited references, polytropes are no strangers to readers of this journal. 

 Then in section V we will answer our question \emph{How much surcharge can a failed white dwarf hold?} non-relativistically.
 We explain why the answer is independent of $\hbar$ even though quantum-mechanical degeneracy pressures hold
up the star against collapse.

 In section VI we explain why the answer, which is also independent of the speed of light $c$,
remains valid when the quantum-mechanical degeneracy pressure-density relation is calculated special relativistically.

 Section VII illustrates our findings.
 
 The conclusions are presented in section VIII, where we also comment on the general relativistic problem.
 
 We use Gaussian units where $4\pi\veps_0^{}=1$.
 To switch to SI units, replace $e^2$ by $e^2/4\pi\veps_0^{}$ everywhere.
\vspace{-15pt}

\section{The Newtonian answer}\vspace{-10pt}

 {In Newtonian physics the lowest energy state of an electron and a proton is achieved when both particles have 
zero kinetic energy and occupy the same location. 
 It is not important for this argument that in Newtonian physics the lowest energy is negative infinite 
--- what matters is that any other configuration of the two particles has a finite  {(hence higher)}
 potential energy, and therefore is not a ground state.
 Likewise, any non-zero kinetic energy only adds a positive amount to the total energy.
 It is understood that any discussion of kinetic energy refers to a center-of-mass inertial frame.}

 We now show that even with many point protons and point electrons, as long as the ratio $\Nneg/\Npos$ falls
within a certain range, the lowest energy state in Newtonian physics is achieved when all particles are at rest
at the same location. 
 The argument is truly elementary and can be presented in a general physics course where gravitational and electrical 
forces are covered.

 Assume first there are $\Npos$ protons but more electrons. 
 Then $\Npos$ disjoint electron-proton pairs can be formed, and, as already noted, 
each pair's ground state is achieved when the particles occupy the same location. 
 The two charges in each such pair neutralize each other, yet different pairs attract each other by Newtonian gravity.
 The mutual potential energy of two such neutral pairs is the lowest when the two pairs occupy the same location. 
 This argument can now be repeated until all neutral pairs occupy the same location.

 Next, suppose that the excess electrons are at first infinitely far removed from this 
neutral completely collapsed configuration of $\Npos$ electron-proton pairs, and also from each other. 
 With the usual convention that potential Coulomb or Newton energy vanishes when two such particles are infinitely far
removed from each other, we now ask whether the potential energy of the system can be lowered by placing electrons 
closer to the location of the already existing collapsed configuration, one after another. 
 The potential energy of a pair of effective point particles a distance $r$ apart, one an electron, 
the other a collapsed configuration of $\Npos$ protons and $\Nneg$ electrons, is 
$[-G(\Npos\mPR +\Nneg\mEL)\mEL -e^2(\Npos-\Nneg)]/r$, and  as long as this expression 
{is $\leq 0$ we can add an extra electron to the collapsed point configuration
without increasing the potential energy when reducing $r$ to zero, each time increasing $\Nneg$ by 1. 
 This yields the upper bound 
\vspace{-.2truecm}
\begin{equation}
\frac{\Nneg}{\Npos}
\leq
\frac{1 + \tfrac{G\mPR\mEL}{e^2}} {1 - \tfrac{G\mEL^2}{e^2}}.
\vspace{-.3truecm}
 \label{eq:NnegOverNposUPPERbound}
\end{equation}}

 Similarly, if there is an excess of protons, one can move yet another proton 
from infinitely far away to the collapsed configuration without increasing the potential energy 
as long as $-G(\Npos\mPR +\Nneg\mEL)\mPR \leq e^2(\Nneg-\Npos)$,  and this translates into the lower bound 
\vspace{-.2truecm}
\begin{equation}
\frac{\Nneg} {\Npos}
\geq
\frac{1 - \tfrac{G\mPR^2}{e^2}}{1 + \tfrac{G\mPR\mEL}{e^2}} .
\vspace{-.3truecm}
 \label{eq:NnegOverNposLOWERbound}
\end{equation}

   {Both bounds are reproduced with non- and special-relativistic quantum-mechanical models, 
and presumably also when general relativity is taken into account.
 However, at this point there is no reason to expect that this should be true, for a stellar ground state with 
between $1.5\cdot 10^{55}$ and $9\cdot 10^{55}$ protons and roughly as many electrons is not collapsed to a point, 
but spread out, and its structure depends on quantum mechanics and relativity.
 Quantum mechanics,  especially the Pauli principle, accounts for the stabilization against collapse; 
cf. \cite{LiebSeiringer}.}
 {It creates an effective \emph{repulsion} among fermions of the same kind.
 This repulsion competes against the overall Newtonian \emph{attraction} among the particles which acts as long as
the above stated bounds on $\Nneg/\Npos$ 
are obeyed, so one would expect tighter bounds than (\ref{eq:NnegOverNposUPPERbound})
and (\ref{eq:NnegOverNposLOWERbound}) to show up quantum-mechanically, featuring $\hbar$.
 Relativity theory destabilizes (Chandrasekhar's critical mass in special, black holes in general relativity), 
so also $c$ should show up in the bounds when relativity is taken into account.}\vspace{-.5truecm}

\section{The role of quantum mechanics}\vspace{-10pt}

  {Physics students early on learn that quantum mechanics prevents the electron and proton
in a hydrogen atom from coalescing onto a point.
 Quantum mechanics also prevents the collapse of a system of many electrons and protons interacting electrically
and gravitationally.} 
 {And the question we raise in this paper takes a precise form}: 
\smallskip

\noindent
   \emph{{Given the number $\Npos$ of protons, for which numbers $\Nneg$ of electrons does a ground state 
of the Hamiltonian exist?}}

 The Hamiltonian $H$ is obtained from the familiar expression $E=E^{\mbox{\tiny{kin}}}+E^{\mbox{\tiny{pot}}}$
of the classical non-relativistic total energy of a system of
$\Npos$ point protons and $\Nneg$ point electrons, interacting pairwise with electrical Coulomb and gravitational
Newton forces, by replacing the classical momentum vector $\pV_k^\pm$ of the $k$-th positively or negatively charged particle 
by $-i\hbar \nabla_k^\pm$ in the kinetic energy terms; here, $\nabla_k^\pm$ 
is the gradient operator with respect to the $k$-th position {vector $\qV_k^\pm$} of the pertinent particle species. 

 Our quantum-mechanical ground state is a normalized energy-mini\-mizing wave function $\Psi$ of $H$,
 with $\Npos$ in the range $1.5\cdot 10^{55}< \Npos< 9\cdot 10^{55}$ and $\Nneg$ nearby.
 ``Normalized'' means $\langle\Psi|\Psi\rangle =1$ in Dirac's notation of the Hilbert space
inner product. 
 ``Energy-minimizing'' means that $\Psi$ is an eigenfunction of $H$, viz.
\begin{equation}
H\Psi = E_g^{}\Psi,
\end{equation}
where $E_g^{}$ is the lowest eigenvalue of $H$.

 $H$ is separately invariant under permutations of the position variables of the electrons and 
of the protons among themselves, respectively.
 Without the Pauli principle, which plays no role in the hydrogen problem,
the ground state can be shown \cite{LiebSeiringer} to feature the same permutation symmetry as $H$,
hence is bosonic.
 Incidentally, because of the dark matter puzzle,
research devoted to the study of hypothetical bosonic stars has attracted attention in recent years, 
e.g. \cite{Juerg}, \cite{KieBOSONS}, and references therein.
 But protons and electrons are not bosons, of course, so
a bosonic ground state has a much too low energy and is spatially too compactified \cite{LiebSeiringer}.
 We now give a simple explanation how the Pauli principle for fermions 
forces the ground state of a failed white dwarf to be much more spread out than that of a bosonic star.

 If the wave function $\Psi$ is restricted to be a many-body Pauli spinor, a complex-valued function of (here)
 $\Npos+\Nneg$ variables $(\qV_k^\pm,\sigma_k^\pm)$, where 
$\sigma_k^\pm$, taking values $-1$ or $1$, is the discrete spin variable of the $k$-th positive or negative particle,
indicating its spin up or down, then the Pauli principle for fermions demands that $\Psi$ changes sign whenever 
the variables of two different protons or two different electrons are permuted, respectively.
 This defines a very small subset of all wave functions, and the lowest eigenvalue which $H$ can 
accomplish on this small set is actually a very very high eigenvalue of the same operator $H$ on the unrestricted 
set of wave functions.
 As students learn from the explicit solution of the hydrogen problem, the higher the eigenvalue the more 
spread out the pertinent eigen wave function is.
 This is also true for the many body problem.
 In a nutshell, it explains the spread out nature of our stellar ground state.

  Of course, nobody is able to solve this problem exactly, unless $\Npos=1=\Nneg$ (the hydrogen problem).
 Also the question how much $\Nneg$ may differ from $\Npos$ for a ground state to exist is open.
 Fortunately, $10^{55}$ is a huge number, as good as $\infty$ for many practical purposes, and this suggests
that an accurate systematic evaluation of the problem can be accomplished by an expansion in descending powers
of the number of particles, similar to the expansion $E_g^{}(N) = C_{7}N^{7/3} + C_{6}N^{6/3} + C_{5}N^{5/3} +\cdots$ 
for an atom with large $N=Z$, which terminates with an error term that is not a simple integer power of $N^{1/3}$; cf.  \cite{FS}.
 The coefficients $C_7$, $C_6$, etc. are independent of $N$ and can be computed in practice.
 In a two-species model with Coulomb and Newton interactions a similar expansion involves $\Npos$ and $\Nneg$. 
 The leading $N^{7/3}$ term can be computed with a Thomas--Fermi model (cf. \cite{Thirring}, chpt.4.2), discussed in this paper.

  In the Thomas--Fermi model all physical quantities, e.g. the ground
state energy $E_g(\Nneg,\Npos)$, are computed in terms of integrals over expressions involving
the particle densities $\nupos(\sV)$ and $\nuneg(\sV)$, where $\sV$ denotes a point in physical space. 
 The particle densities are defined as the quantum-mechanical expectation value $\langle \Psi | \Delta^\pm(\sV)|\Psi\rangle$ 
of the microscopic particle densities $\Delta^\pm(\sV)=\sum_k\delta(\sV-\qV_k^\pm)$, which
for huge numbers $\Npos$ and $\Nneg$ are (by the law of large numbers)
close to the actual particle densities \cite{Thirring} ---
which can be computed  to leading order in the expansion via the following system of differential equations.\vspace{-10pt}
 
\section{The Thomas--Fermi model of \\ a failed white dwarf star}

 The Thomas--Fermi equations of structure for a non-rotating, {zero-temperature}
white dwarf star composed of electrons and various types of nuclei can be found in Chandrasekhar's
classic book \cite{Chandra}, also in the book \cite{KippenhahnWeigert},  and in \cite{SilbarReddy}, \cite{Garfinkle}, 
and \cite{KNY}.
  For a non-rotating star one may assume spherical symmetry, so all
the basic structure functions are then functions only of the radial distance $r$ from the star's center,
and the differential equations involved in the discussion reduce to the ordinary type.

 We simplify the model in order to describe a failed star composed only of protons and electrons,
both of which are spin-$\frac12$ fermions.
 Each species is treated as an ideal quantum gas in its own right.
 The number density functions $\nupos(r)\geq 0$ and $\nuneg(r)\geq 0$ are
assumed to integrate to the total number of protons, respectively electrons, viz.
\begin{eqnarray}
\int_{\Rset^3} \nu_f^{}(r) d^3r = N_f^{},
 \label{eq:Nf}
\end{eqnarray}
where  ${}_f={}_p$ or ${}_f={}_e$.
 The protons have rest mass $\mPR$ and charge $+e$; the electrons have rest mass $\mEL$ and charge $-e$.
 Thus the mass density of the star is given by
\begin{equation}
\mu(r) = \mPR \nupos(r) + \mEL\nuneg(r)
 \label{eq:massdensity}
\end{equation}
and its charge density by
\begin{equation}
\sigma(r) = e \nupos(r) - e \nuneg(r).
 \label{eq:chargedensity}
\end{equation}
 This means the star is overall neutral if $\Npos=\Nneg$; otherwise it carries a surcharge which may have
either sign.

 The electrons and protons jointly produce a Newtonian gravitational potential $\phi_N^{}(r)$ and 
an electric Coulomb potential $\phi_C^{}(r)$.
 The Newton potential $\phi_N^{}$ is related to the mass density $\mu$ by a radial Poisson equation,
\begin{equation}
\left(r^2\phi_N^{\prime}(r)\right)^\prime
  = 4\pi G \mu(r)r^2,
 \label{eq:PoissonN}
\end{equation}
where $G$ is Newton's constant of universal gravitation.
 Similarly, the Coulomb  potential $\phi_C^{}$ is related to the charge density $\sigma$ by a radial Poisson equation,
\begin{equation}
-\left(r^2\phi_C^{\prime}(r)\right)^\prime = 4\pi \sigma(r)r^2.
 \label{eq:PoissonC}
\end{equation}
 As usual, the primes in (\ref{eq:PoissonN}) and (\ref{eq:PoissonC})
mean derivative with respect to the displayed argument, in this case $r$.

 Each species, the electrons and the protons, satisfies an Euler-type mechanical force balance equation,
\begin{eqnarray}
\nupos(r)\left[-\mPR \phi_N^{\prime}(r) - e \phi_C^{\prime}(r)\right] -  \ppos^\prime(r) =0,
 \label{eq:forceBALANCEpos}\\
\nuneg(r)\left[-\mEL \phi_N^{\prime}(r) + e \phi_C^{\prime}(r)\right] -  \pneg^\prime(r) =0.\;
 \label{eq:forceBALANCEneg}
\end{eqnarray}
 Here, $\ppos$ and $\pneg$ are the degeneracy pressures of the ideal proton and electron gases (at zero Kelvin).
 They are
computed in any introductory statistical mechanics course which covers ideal quantum gases, e.g. 
\cite{Balian}, \cite{Balescu}
and can also be found in \cite{SilbarReddy} and \cite{Garfinkle}.
 For a non-relativistic gas of spin-$\frac12$ fermions (subscript ${}_f$) of mass $m_{\text{f}}^{}$ and number 
density $\nu_{\text{f}}^{}$ one has (see, e.g. \cite{Chandra}, p.362; see also \cite{Norm})
\begin{equation}
p_{\text{f}}(r) = \frac{\hbar^2}{m_{\text{f}}}\frac{(3\pi^2)^{2/3}}{5}\nu^{5/3}_{\text{f}}(r);
 \label{eq:pDEG}
\end{equation}
here, ${}_f$ $=$ ${}_p$ or ${}_e$, and $\hbar$ is the reduced Planck constant.

 For later reference, we recall that any pressure-density relation 
of the type $p = K_\gamma\nu^\gamma$ for some constant $K_\gamma$ is
called a \emph{polytropic} law of power $\gamma$,  associated with a polytropic index $n=1/(\gamma-1)$.
 So for the fermionic degeneracy pressure, $\gamma=5/3$ and $n=3/2$.

 The system of Thomas--Fermi equations can be reduced to a closed system of equations for the densities 
$\nupos(r)$ and $\nuneg(r)$ alone.
{We level the ground for this by noting that one can show that the two density 
functions are monotonic decreasing from their central value to zero. 
 The smallest $r$ value for which $\nupos(r)=0$ is denoted by $\rpos$; similarly we define $\rneg$ 
in terms of $\nuneg(r)$.
 We stipulate that $\nupos(r)=0$ for all $r>\rpos$, and similarly $\nuneg(r)=0$ for all $r>\rneg$.
 Clearly, either $\rneg< \rpos$, or $\rneg = \rpos$, or $\rneg> \rpos$.
 Accordingly we distinguish three possible regions, called}

(a) the bulk region where $\nupos(r)>0$ and $\nuneg(r)>0$,

(b) the positive atmosphere: $\nupos(r)>0$ and $\nuneg(r)=0$,

(c) the negative atmosphere: $\nupos(r)=0$ and $\nuneg(r)>0$.

\noindent
 A pair of densities $(\nupos(r),\nuneg(r))$ may exhibit only the bulk region (namely when $\rpos=\rneg$),
but generally it exhibits either the combination (a) \& (b) or (a) \& (c), depending on whether 
$\rpos>\rneg$ or $\rneg>\rpos$, respectively.
 Incidentally, what we call ``atmosphere'' should not be confused with the stellar atmospheres 
that are observed and studied by astronomers.
 It is simply a name for an outer low-density region where either one or the other density has vanished, but not both.

 Beginning with the bulk region, since both $\nupos>0$ and $\nuneg>0$, both 
Eqs.(\ref{eq:forceBALANCEpos}) and (\ref{eq:forceBALANCEneg}) are nontrivial. 
 Therefore,

1) use (\ref{eq:massdensity}) and (\ref{eq:chargedensity}) to express $\mu$ and $\sigma$ in terms of 
$\nupos$ 

\quad and $\nuneg$ in (\ref{eq:PoissonN}) and (\ref{eq:PoissonC});

2) next multiply (\ref{eq:PoissonN}) by $-\mPR$ and (\ref{eq:PoissonC}) by $e$ and add the

\quad resulting two equations, then use (\ref{eq:forceBALANCEpos}) to replace

\quad $-\mPR \phi_N^{\prime}(r) - e \phi_C^{\prime}(r)$ in terms of $\nupos(r)$ 
and $\ppos^\prime(r)$;

3) next, use (\ref{eq:pDEG}) to express $\ppos$ in terms of $\nupos$;

4) similarly, multiply (\ref{eq:PoissonN}) by $-\mEL$ and (\ref{eq:PoissonC}) by $-e$ and 

\quad also add these equations, then use (\ref{eq:forceBALANCEneg}) to replace 

\quad $-\mEL \phi_N^{\prime}(r) + e \phi_C^{\prime}(r)$ in terms of 
$\nuneg(r)$ and $\pneg^\prime(r)$;

5) next use (\ref{eq:pDEG}) to express $\pneg$ in terms of $\nuneg$.
 
6) Now use simple algebraic and calculus manipulations to obtain the following system of nonlinear second-order 
differential equations for the density functions $\nupos$ and $\nuneg$,
valid wherever both  $\nupos(r)>0$ and $\nuneg(r)>0$:
\begin{widetext}
\begin{eqnarray}
- \veps \zeta \frac{1}{r^2}\Ddr\left(r^2\Ddr \nupos^{2/3}(r)\right)
&\,\  =  - \left(1 -\frac{G\mPR^2}{e^2}\right) \nupos(r) + \left(1 +\frac{G\mPR\mEL}{e^2}\right) \nuneg(r),
 \label{eq:PoissonMUpTHREEhalf}\\
- \zeta \frac{1}{r^2}\Ddr\left(r^2\Ddr \nuneg^{2/3}(r)\right)
&\!\!  = \left(1 + \frac{G\mPR\mEL}{e^2}\right) \nupos(r) - \left(1 - \frac{G\mEL^2}{e^2}\right) \nuneg(r).
 \label{eq:PoissonMUeTHREEhalf}
\end{eqnarray}
\end{widetext}
 Here, $\veps:= {\mEL}/{\mPR}$ and  
$\zeta:= ({3^{2/3}\pi^{1/3}}/{8}){\hbar^2}/{\mEL e^2}$.
 Steps 1)--6) could also be assigned to a student for practice.

 Coming to the atmospheric regions, a positive atmosphere is governed by
(\ref{eq:PoissonMUpTHREEhalf}) with $\nuneg(r)=0$, while a negative atmosphere is governed by
(\ref{eq:PoissonMUeTHREEhalf}) with $\nupos(r)=0$. 

 Each equation is of second order and requires two initial conditions.
 At the bulk-atmosphere interface, located either at $\rpos$ or $\rneg$ (whichever is smaller),
the density of the species which forms the atmosphere needs to be continuously differentiable.
 In the bulk, conditions are posed at $r=0$. 
 Naturally $\nupos^\prime(0)=0=\nuneg^\prime(0)$. 
 The values of $\nupos(0)$ and $\nuneg(0)$ are to be chosen such that (\ref{eq:Nf}) holds.
 This will not be possible for arbitrary combinations of $\Nneg$ and $\Npos$.
 In particular, we will show that only a finite interval of ratios $\Nneg/\Npos$ will be allowed.

 This system of coupled differential equations for the density functions $\nupos$ and $\nuneg$ in bulk and
atmosphere regions generalizes the single Lane--Emden equation for the polytrope of index $n=\frac32$, 
which has only a bulk interior; see \cite{Chandra},  \cite{KippenhahnWeigert}, \cite{SilbarReddy}, \cite{Garfinkle}.

 As for the numerical values of the parameters,
$\veps \approx 1/1836\approx 5.54\cdot 10^{-4}$ and  
$\zeta\approx 52.185\frac{\hbar}{\mEL c}$, where $\frac{\hbar}{\mEL c}\approx 3.86\cdot 10^{-13}$m
is the electron's reduced Compton wave length. 
 The three ratios of gravitational-to-electrical coupling constants that appear in the coefficient matrix at the
right-hand sides of Eqs.(\ref{eq:PoissonMUpTHREEhalf}) and (\ref{eq:PoissonMUeTHREEhalf}) are fantastically tiny numbers, viz.
  ${G\mEL^2}/{e^2}\approx 2.400\cdot 10^{-43}$, ${G\mPR\mEL}/{e^2}\approx 4.407\cdot 10^{-40}$, and
${G\mPR^2}/{e^2}\approx 8.09\cdot 10^{-37}$.
 Yet one has to resist the impulse to neglect these tiny numbers versus 1 in the cofficients, for
this would result in a singular coefficient matrix, and there would not be any nontrivial solution pair $(\nupos,\nuneg)$.
{This is intuitively clear because
the three tiny ratios of coupling constants are the only places featuring Newton's constant of universal
gravitation, $G$, and it is gravity, not electricity, that binds the ideal Fermi gases together to form a star.}

 The nonlinearity of Eqs.(\ref{eq:PoissonMUpTHREEhalf}) and (\ref{eq:PoissonMUeTHREEhalf}), coupled to each other and to
their atmospheric counterparts, stands in the way of solving them generally in closed form. 
  Yet we can obtain a sufficient amount of insight into their solutions to allow us conclusions about the 
interval of $\Nneg/\Npos$ values for which solutions exist. \vspace{-10pt}

\section{Extremal stellar surcharges}\label{ex}\vspace{-10pt}

 As explained in the introduction, our quest for the extremal surcharge on an object
implicitly assumes that the object's number of nuclei is considered fixed 
when the number $\Nneg$ of electrons is varied.
 Thus, our question for the stellar surcharge can be rephrased: 
\emph{How many electrons per proton can a failed white dwarf bind?}

 Recall that the traditional single-density models are based on the local neutrality approximation, which
implies $\Npos=\Nneg$, and they are thought to
yield a very good approximation to the overall mass density function (not the charge density, of course).
 So one should expect that a non-neutral pair $(\Npos,\Nneg)$ will correspond to a ratio $\Nneg/\Npos\approx 1$. 
 We will make this quantitatively precise.

 Consider first a star with negative atmosphere, not necessarily the extreme case.
 We multiply Eq.(\ref{eq:PoissonMUeTHREEhalf}) by $4\pi r^2$ and integrate over $r$ 
from $0$ to the distance $\rneg$ where the negative atmosphere vanishes.
 (Strictly speaking, (\ref{eq:PoissonMUeTHREEhalf}) is a-priori only valid inside the bulk region, but 
we can extend (\ref{eq:PoissonMUeTHREEhalf}) to all $r$ by noting that $\nupos(r)=0$ for $r\geq \rpos$ in this case.)
{Using $4\pi \int_0^{\rneg} \nupos(r)r^2 dr =\Npos$ and $4\pi \int_0^{\rneg} \nuneg(r)r^2dr =\Nneg$
at the right-hand side, and the fundamental theorem of calculus on the left, 
we obtain
\begin{eqnarray}
\hspace{-15pt}
4\pi \zeta \rneg^2 {\nuneg^{\frac23}}{}^\prime(\rneg)
 =  \left(1 - \tfrac{G\mEL^2}{e^2}\right) \Nneg -\left(1 + \tfrac{G\mPR\mEL}{e^2}\right) \Npos .
 \label{eq:PoissonMUeINT}
\end{eqnarray}
 But $\nuneg^{\prime}(\rneg)$ is the left-derivative of $\nuneg(r)$ at the radial distance $\rneg$ where 
the density $\nuneg(r)$ reaches $0$, and since an otherwise positive function cannot reach $0$ with a positive slope, 
{it follows that $\nuneg^{\prime}(\rneg)\leq 0$, so ${\nuneg^{\frac23}}{}^\prime(\rneg)\leq 0$, 
and so (\ref{eq:PoissonMUeINT}) implies (\ref{eq:NnegOverNposUPPERbound}).}

 The case of a positive atmosphere is treated in a completely analogous manner to obtain 
\begin{equation}
\hspace{-2pt}
4\pi\veps \zeta 
 \rpos^2{\nupos^{\frac23}}{}^\prime(\rpos)
 = \left(1 - \tfrac{G\mPR^2}{e^2}\right)\!\Npos - \left(1 + \tfrac{G\mPR\mEL}{e^2}\right)\! \Nneg,\hspace{-4pt}
 \label{eq:PoissonMUpINT}
\end{equation}
{and from (\ref{eq:PoissonMUpINT}) one deduces (\ref{eq:NnegOverNposLOWERbound}).
 This part can also be assigned to students as an exercise.}}

  {Since our argument only involves the fact that $r_f^2{\nu_f^{\frac23}}{}^\prime(r_f^{})$ cannot be $>0$, 
where ${}_f={}_p$ or ${}_f={}_e$, 
at this point we do not yet know whether (\ref{eq:NnegOverNposUPPERbound}) and (\ref{eq:NnegOverNposLOWERbound}) 
are sharp bounds for the Thomas--Fermi model.
 We recall that in the effective single-density $n=\frac32$ polytrope model of \cite{Chandra}
the solution becomes zero  at a finite radius, with a strictly negative slope,
and if the same conclusion would hold also for \emph{all} solutions of our system of $n=\frac32$ equations,
 then the left-hand sides of (\ref{eq:PoissonMUeINT}) and (\ref{eq:PoissonMUpINT}) 
would be {strictly negative and feature $\hbar$}, in particular for extremely charged ones. 
 Interestingly, this is only true for \emph{almost all} solutions --- 
there are two exceptional cases, and these are exactly the two extremely surcharged solutions!}

 Indeed, it can be shown that two extreme atmosphere solutions exist which satisfy
\begin{equation}
r^2 {\nu_f^{\frac23}}^\prime(r^{}) \to 0,
 \label{eq:NnegOverNposSATURATED}
\end{equation}
as $r\rightarrow r_f^{}$, where ${}_f={}_p$ or ${}_f={}_e$, depending on whether the atmosphere consists of
protons or electrons, respectively.
 Coupled with standard uniqueness results which can be found in, e.g. \cite{Boyce}, this
implies that $r_f^{}$ is $\infty$.
 This unexpected result explains, per (\ref{eq:NnegOverNposSATURATED}) 
for ${}_f={}_p$ or ${}_f={}_e$, respectively, 
inserted into (\ref{eq:PoissonMUeINT}) or (\ref{eq:PoissonMUpINT}), 
why (\ref{eq:NnegOverNposUPPERbound}) and (\ref{eq:NnegOverNposLOWERbound}), in which $\hbar$ plays no role, 
are also sharp in the Thomas--Fermi model where $\hbar$ features in $\zeta\propto\hbar^2$, and implicitly
in $\nu_f^{}$, at the left-hand sides of (\ref{eq:PoissonMUeINT}) and (\ref{eq:PoissonMUpINT}).

 {We have of course left out many mathematical details in the previous paragraph; for the interested reader we have 
included {some} details in an extended version of this short pedagogical paper \cite{HK}. 
 In this extended version, we have also included an explicitly solvable approximate model  {\cite{KNY} for the benefit of those 
who contemplate including in an introductory astrophysics course a discussion of
the basic equations of stellar structure, in particular for white dwarfs.
 If, after switching to a dimensionless notation,
one replaces the polytropic power $5/3$ of the pressure-density relation, predicted by non-relativistic 
quantum mechanics, with the nearby $6/3$, corresponding to a polytropic index $n=1$, one obtains this 
{toy} model of a failed white dwarf star which captures the essence of the $5/3$ model qualitatively correctly.}
 This model produces exactly the same surcharge bounds, is solvable in terms of} {simple
elementary functions, and features a family of two-species solutions, 
almost all of which are of finite extent, except for two extremely surcharged solutions.}

  We may thus summarize our discussion with stating the allowed range of ratios $\Nneg/\Npos$ in a single formula,
\begin{eqnarray}
\boxed{
\frac{1 -\tfrac{G\mPR^2}{e^2}}{1 +\tfrac{G\mPR\mEL}{e^2}}
\leq 
\frac{\Nneg}{\Npos}
\leq 
\frac{1 + \tfrac{G\mPR\mEL}{e^2}}{1 - \tfrac{G\mEL^2}{e^2}}
      }\;.
 \label{eq:NposToNnegINTERVAL}
\end{eqnarray}
 Formula (\ref{eq:NposToNnegINTERVAL}) is the answer to the question how 
much surcharge a failed white dwarf star can hold, given $\Npos$, {as per the non-relativistic
Thomas--Fermi model.} 

 From (\ref{eq:NposToNnegINTERVAL}) we have, to very good approximation,
\begin{eqnarray} \label{eq:NposToNnegINTERVALexpand}
\left(1 - \tfrac{G\mPR\mEL}{e^2} \right){\Nneg}
\leq 
{\Npos}
\leq 
\left(1 +\tfrac{G\mPR^2}{e^2} \right)\Nneg,
\end{eqnarray}
or
\begin{eqnarray}\label{eq:NposMinusNneg}
- \tfrac{G\mPR\mEL}{e^2}{\Nneg}
\leq 
{\Npos-\Nneg}
\leq 
\tfrac{G\mPR^2}{e^2} {\Nneg}.
\end{eqnarray}
 So a neutral failed white dwarf star with {$\Npos = 5\cdot 10^{55} = \Nneg$}
can be stripped of {$\approx 4\cdot 10^{19}$} electrons, while {$\approx 2\cdot 10^{16}$}
electrons can be deposited on it without changing $\Npos$.\vspace{-15pt}

\section{The ``universal'' validity of the surcharge bounds}\label{u}\vspace{-10pt}

 We have just explained the somewhat unexpected fact that the non-relativistic
Thomas--Fermi model yields the $\hbar$-independent surcharge bounds (\ref{eq:NposToNnegINTERVAL}),
even though the structure of the solutions depends on $\hbar$ through the quantum-mechanical degeneracy 
pressure.
 The fact that the surcharge bounds (\ref{eq:NposToNnegINTERVAL}) do not depend on the speed of light $c$
is of course not surprising, since we have worked with a non-relativistic model.
 We will now present a compelling argument for why our surcharge bounds (\ref{eq:NposToNnegINTERVAL}) are the
correct bounds for a failed white dwarf star also in the special-relativistic regime!
 Gravity remains Newtonian.

 It is clear that the key to (\ref{eq:NposToNnegINTERVAL}) is the observation that the density of
an extremal atmosphere is infinitely extended, and goes to zero with vanishing slope rapidly enough, 
unlike the non-extremal densities. 
 This remains true if we take special relativity into account in the manner done by Chandrasekhar \cite{Chandra}.
 We set $\veps_f=\veps$ if ${}_f={}_p$ and $\veps_f=1$ if ${}_f={}_e$.
 Then in Eqs.(\ref{eq:PoissonMUpTHREEhalf}) and (\ref{eq:PoissonMUeTHREEhalf}) 
one needs to change $\veps_f\zeta\nu^{2/3}_f \to a_f\sqrt{1 + \ell_f^2 \nu_f^{2/3}}$, with
$a_f = \frac{m_f c^2}{4\pi e^2}$ and $\ell_f = (3\pi^2)^{1/3} \hbar/m_fc$.
 Modulo additive and multiplicative constants the relativistic square root expression
interpolates continuously between $\nu^{2/3}_f$ and $\nu^{1/3}_f$,
according to whether $\ell_f \nu_f^{1/3}$ is $\ll 1$ or $\gg 1$, respectively.
 The integration of the counterpart to Eqs.(\ref{eq:PoissonMUpTHREEhalf}), (\ref{eq:PoissonMUeTHREEhalf})
will produce the counterpart to Eqs.(\ref{eq:PoissonMUeINT}) and (\ref{eq:PoissonMUpINT}), and therefore 
(\ref{eq:NposToNnegINTERVAL}).
{Furthermore, as in the non-relativistic setting, one can show that two extremely surcharged solutions exist
which are infinitely extended, for which 
\begin{equation}
r^2\tfrac{d}{dr} \sqrt{1 + \ell_f^2 \nu_f^{2/3}(r)} \to 0,
 \label{eq:NnegOverNfSATURATED}
\end{equation}
as $r\rightarrow r_f^{}=\infty$, where ${}_f={}_p$ or ${}_e$ according to whether the extremal atmosphere is made of 
protons or electrons. 
 So (\ref{eq:NposToNnegINTERVAL}) is again sharp.}
  Thus we have arrived at the perhaps even more unexpected result:

 \emph{The optimal bounds on $\Nneg/\Npos$ for a relativistic failed white dwarf are 
given by (\ref{eq:NposToNnegINTERVAL}), independently of $\hbar$ and $c$}.

 In fact, these bounds are 
actually (largely) independent of the pressure-density relations for the two species. 
 We write ``largely'' because some relations will not lead to solutions 
with finite mass but yield $\Npos=\infty=\Nneg$, rendering the surcharge question pointless.
 An important example are polytropes with index $n>5$.
 Also the isothermal, merely partially degenerate ideal Fermi gas at temperature $T>0$
leads to stars with infinite mass.

\vspace{-20pt}

\section{Illustrations} \vspace{-5pt}
 To illustrate our findings there are two compromises to be made, though.
 Namely, the fantastically tiny ratios of the gravitational to electrical coupling
constants between electron and proton definitely cause headaches, but also the small mass ratio $\mEL/\mPR\approx1/1836$
is a source of trouble. 
 Both these small numbers taken together make it just impossible to produce any useful graphs at all. 

 However, it is only necessary to illustrate the findings qualitatively correctly.
 In this vein, in the following we illustrate our findings for 
the science fiction values $G\mPR^2/e^2=1/2$ and $\mEL/\mPR=1/10(=\veps)$; for consistency, therefore,
$G\mPR\mEL/e^2=\veps/2$ and $G\mEL^2/e^2=\veps^2 /2$.
{The other physical constant of the model, $1/\alphaS \approx 137.036$.}
 
 We begin with the extremely surcharged solutions. \vspace{-.4truecm}

\begin{figure}[ht]
  \includegraphics[width = 8.2truecm,scale=1.1]{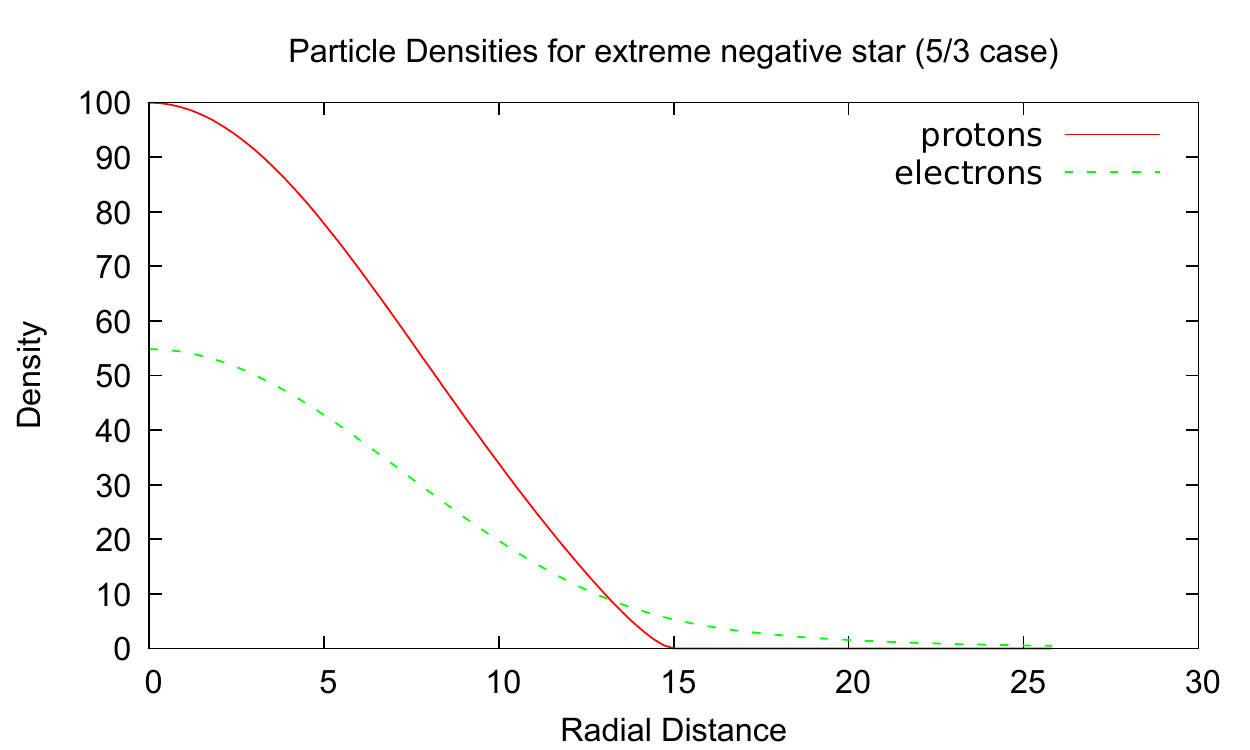} 
\vspace{-.4truecm}
\caption{Shown are the density functions $\nupos(r)$ and $\nuneg(r)$ for the upper
extreme ratio $\Nneg/\Npos = {( 1 + \tfrac{G\mPR\mEL}{e^2})}/{(1 - \tfrac{G\mEL^2}{e^2})}$,
 with science fiction values $G\mPR^2/e^2=1/2$ and $\mEL/\mPR=1/10$.}  \vspace{-1truecm}
\end{figure}
\begin{figure}[ht]
  \includegraphics[width = 8.2truecm,scale=1.1]{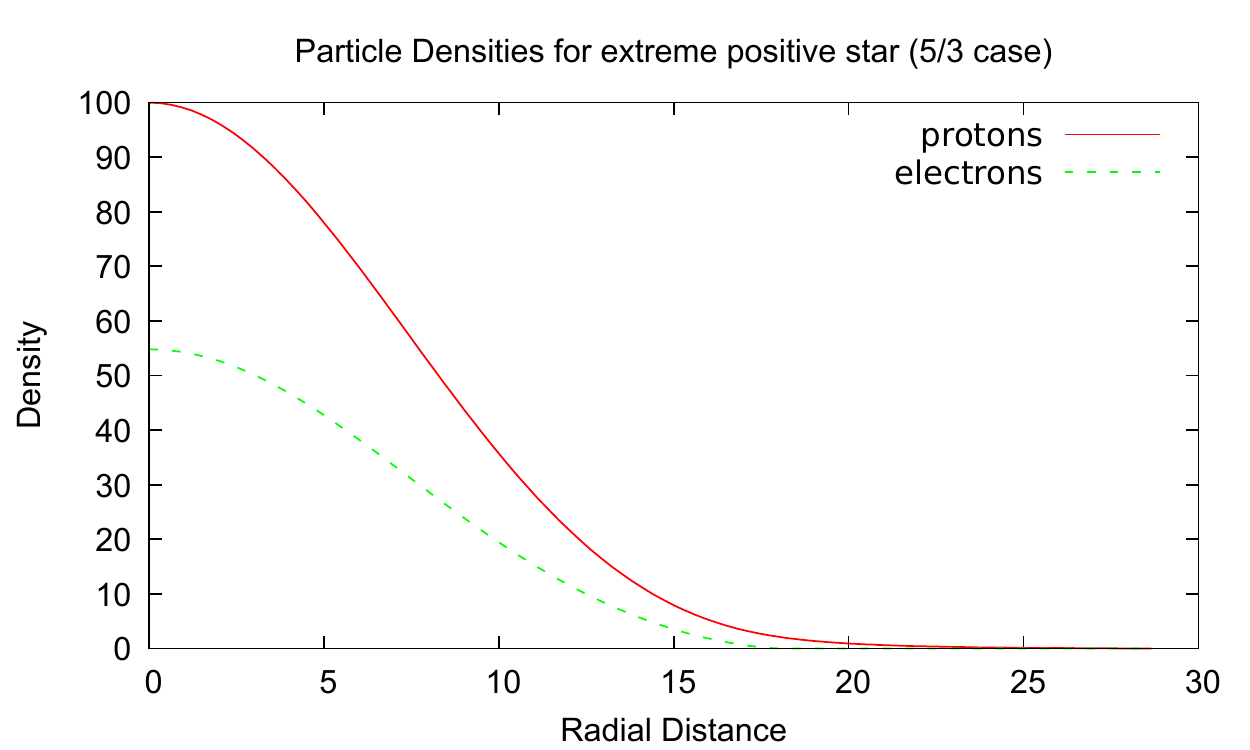} 
\vspace{-.4truecm}
\caption{Shown are the density functions $\nupos(r)$ and $\nuneg(r)$ for the lower extreme 
ratio $\Nneg/\Npos = {(1 - \tfrac{G\mPR^2}{e^2} )}/{( 1 + \tfrac{G\mPR\mEL}{e^2})}$, 
with science fiction values $G\mPR^2/e^2=1/2$ and $\mEL/\mPR=1/10$. }  \vspace{-.4truecm}
\end{figure}

 Figs.~1 \&\ 2 show the particle density functions of the 5/3 model
with the science fiction values given to the physical constants.
 The central proton density is the same in all examples.
 Note that it \emph{always} exceeds the central electron density. 
 This is so because the more massive protons couple more strongly to gravity than do the lighter electrons,
 while both couple equally strongly electrically; in addition, the stabilizing degeneracy pressure is lower for particles
with larger mass.
 This asymmetry causes a non-zero charge density which varies over the same scale as the bulk, then falls off 
to zero in the atmospheric region (whenever there is one). 

 Another distinguished pair of densities are those of a star without atmosphere, 
when both $\nupos(r)$ and $\nuneg(r)$ vanish at the same dimensionless bulk radius, 
i.e. $\rpos=\rneg$; see Fig.~3.
 For the 5/3 model it can be shown that both densities are scaled $n=3/2$  standard polytropes,
as we now explain. 
 This could be another nice exercise for a student. 

 Insert the ansatz $\nuneg(r) = \lambda \nupos(r)$  into 
Eqs.(\ref{eq:PoissonMUpTHREEhalf}) and (\ref{eq:PoissonMUeTHREEhalf}).
 This generally overdetermines the problem, unless the compatibility condition 
$A\eta^5 +B\eta^3 +C\eta^2+D =0$ is fulfilled, where $\eta :=\lambda^{1/3}$,
and  $A= \left(1+ {G\mEL\mPR}/{e^2}\right)/\veps>0$, 
$B= \left(1 - {G\mEL^2}/{e^2}\right)>0$, $C = -\left(1 -{G\mPR^2}/{e^2}\right)/\veps <0$, 
and $D= - \left(1+ {G\mEL\mPR}/{e^2}\right)<0$.
 Alas, there generally does not exist a solution in closed form, but 
from the signs of the coefficients in this polynomial one can deduce right away that there exists a unique
positive solution $\eta_+$, say, which is near 1, and for $\lambda=\eta_+^3$
both (\ref{eq:PoissonMUpTHREEhalf}) and (\ref{eq:PoissonMUeTHREEhalf}) reduce to the equation 
\begin{eqnarray}
\hspace{-15pt}
\veps \zeta \frac{1}{r^2}\!\!\left(r^2\nupos^{\frac23}{}^\prime(r)\right)^\prime\!\!
= \! \left[\!1 -\tfrac{G\mPR^2}{e^2} -\lambda\! \left(\!1\! +\!\tfrac{G\mPR\mEL}{e^2}\right)\! \right]\!
\nupos(r)\!,
 \label{eq:PoissonMUpMUe}
\end{eqnarray}
which is equivalent (not identical) to the polytropic equation of index $n=3/2$.
 Indeed, setting $\nupos^{\frac23}(r)=C_1\theta(\xi)$ and rescaling $r=C_2\xi$ appropriately
converts (\ref{eq:PoissonMUpMUe}) into the standardized format 
$-\frac{1}{\xi^2}\left(\xi^2\theta^{\prime}(\xi)\right)^\prime = \theta_+^{3/2}(\xi)$, cf. \cite{Emden}, \cite{Chandra},
and thus the no-atmosphere densities are obtained by rescaling the standardized $n=3/2$ polytrope.\vspace{-.6truecm}

\begin{figure}[ht]
  \includegraphics[width = 8.2truecm,scale=1.1]{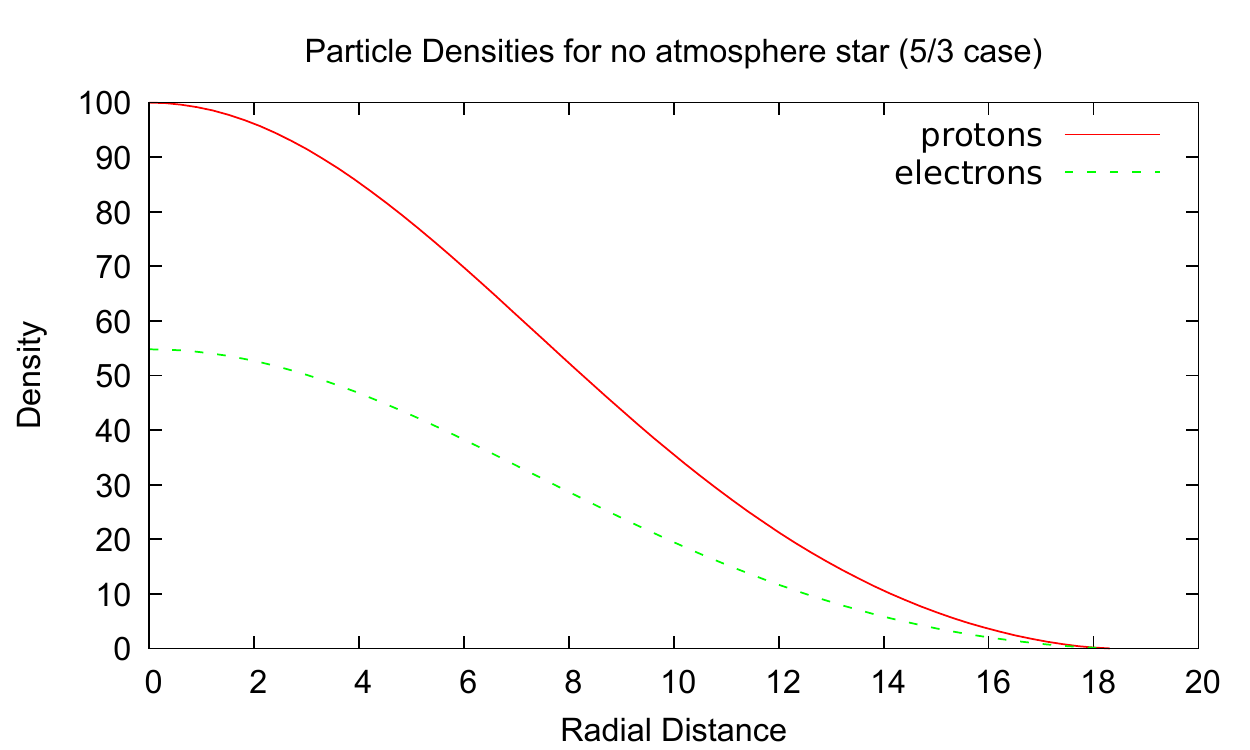} 
 \vspace{-.2truecm}
\caption{Shown are the density functions $\nupos(r)$ and $\nuneg(r)$ for a star without
atmosphere, for science fiction values $G\mPR^2/e^2=1/2$ and $\mEL/\mPR=1/10$. } \vspace{-.2truecm}
\end{figure}
 
{The number of electrons per proton of this no-atmosphere solution is easily computed, also by a student, 
through multiplying Eqs.(\ref{eq:PoissonMUpTHREEhalf}) and (\ref{eq:PoissonMUeTHREEhalf}) with $4\pi r^2$
and integrating.}
 With $\lambda=\eta_+^{3}$ one finds
\begin{equation}
\frac{\Nneg}{\Npos}
 = 
\frac{\lambda^{2/3}\left(1 -\frac{G\mPR^2}{e^2}\right)+ \veps\left(1 +\tfrac{G\mPR\mEL}{e^2}\right)}
{\lambda^{2/3}\left(1 +\tfrac{G\mPR\mEL}{e^2}\right)+ \veps\left(1 -\frac{G\mEL^2}{e^2}\right)}\approx 1.
 \label{eq:NnegOverNposNOatmoEXPLICITfor5third}
\end{equation}
 \vspace{-25pt}

	\section{Conclusions}\label{sec:CONCLUSIONS}\vspace{-10pt}

 We have discussed an {idealized} non-relativistic model of a failed white dwarf star
consisting of electrons and protons, with polytropic power $\gamma=5/3$ of the pressure-density 
relations, predicted by non-relativistic quantum mechanics in the large $N$ limit.
 We also briefly addressed the changes needed when implementing Chandrasekhar's
special-relativistic pressure-density relation which interpolates between a polytropic $\gamma=5/3$
and a $\gamma=4/3$ law.
 We have explained that the number of electrons per proton, $\Nneg/\Npos$, of a solution always 
lies in the interval (\ref{eq:NposToNnegINTERVAL}), no matter which model is used. 
 (In an expanded version \cite{HK} of this short paper we completely solve an exactly solvable
model {\cite{KNY}} obtained by replacing the polytropic $\gamma =5/3$ with the nearby $\gamma = 6/3$, and which 
reproduces the interval (\ref{eq:NposToNnegINTERVAL}) by explicit computation.
 This model could not be incorporated here due to length restrictions.)

 We suspect that (\ref{eq:NposToNnegINTERVAL}) also holds general relativistically, 
which requires the discussion of the Einstein field equations
coupled with both the matter equations for the Fermi gases and the Maxwell equations of the electrostatic
field in curved spacetime \cite{RRb}.
 Yet note that the key observation in our derivation of (\ref{eq:NposToNnegINTERVAL}) is the behavior of
the atmospheric densities at spatial infinity, and in an asymptotically flat spacetime this is the 
region where the general relativistic equations are expected to go over into the non- or special-relativistic equations
discussed here --- hence our conjecture that (\ref{eq:NposToNnegINTERVAL}) 
is valid also in the general relativistic model.

 However, our derivation is only valid for idealized failed stars made of protons and electrons,
with $\Npos$ between about $1.5\cdot 10^{55}$ to about $9\cdot 10^{55}$,
far away from the Chandrasekhar critical mass $C(\Nneg/\Nnuc)^2(\hbar c/G)^\frac32/\mPR^2$,
beyond which no stellar equilibrium is possible, as per special relativity theory;
here, $\Nnuc$ is the number of nucleons in the star and $C$ a numerical factor (see \cite{Chandra}, \cite{Garfinkle}).
  {A more realistic model of a failed white dwarf should use $\approx 92\%$ protons and
$\approx 8\%$ $\alpha$ particles, according to the primordial nucleosynthesis retrodictions of the standard
model of cosmology, necessitating a three-species model of which only two are fermions, the third one being bosons. 
 Instead of a system of Thomas--Fermi equations, one then needs a Thomas--Fermi--Hartree system of equations.
 This may introduce Planck's quantum $h$ into the extremal surcharge bounds, and
having more than two species of particles will be reflected in the surcharge bounds anyhow.}
 Also, near the critical mass for a white dwarf 
the surcharge bounds may change
non-analytically into something more complicated, inheriting its dependence on $\hbar$ and $c$.
 
 Back to formula (\ref{eq:NposToNnegINTERVAL}) for the allowed ratios $\Nneg/\Npos$ in an electron-proton model.
 It is very elementary, and so is its derivation from Newtonian considerations 
as discussed in section II; also its derivation from the Thomas--Fermi equations 
as discussed in sections \ref{ex} and \ref{u} is quite straightforward. 
 This and its surprising universality, i.e. its independence of $\hbar$ and $c$ even in 
a relativistic quantum-mechanical context of ideal Fermi gases, should make it of interest for various types of 
physics courses in which the issue of stability of matter under gravity and electricity can
be raised, cf. \cite{LiebSeiringer}.
 
 A rigorous vindication, in the spirit of \cite{LY}, of our claim that in the Thomas--Fermi models 
it can be shown that the density of an extreme atmosphere goes sufficiently 
rapidly to zero together with its derivative when $r\to\infty$, will be published by the first author 
in a mathematical journal.
 Such an advanced analysis is not suitable for a general physics course, which is why
we only remarked that ``it can be shown.'' 
 Yet, as a compromise, one can verify the analogous claim, and with it (\ref{eq:NposToNnegINTERVAL}),
 by explicit calculation at least for the $6/3$-model, as done in \cite{HK}.

\section*{Acknowledgment}  \vspace{-.2truecm} 
 We thank Elliott H. Lieb for interesting discussions and encouragement. 
 We also thank the referees and the editors for their helpful suggestions.
 After this paper was submitted and the extended version \cite{HK} posted online, Andrey Yudin kindly
pointed out to us that the exactly solvable 6/3 model is also discussed in section IV.B. of \cite{KNY};
the optimal surcharge bounds are not in \cite{KNY}, though. 
 We thank him for his communications, in which he also remarked that test charges are 
attracted to a finite-radius charged star as long as (\ref{eq:NnegOverNposUPPERbound}) and (\ref{eq:NnegOverNposLOWERbound}) 
hold in strict form. 
 We also were looking forward to receiving comments from Norman E. Frankel, which he would offer in his inimitable style. 
 Sadly, Norm passed away while we were readying our paper for submission.

\section*{References}

\bibliographystyle{unsrt}

\end{document}